\documentclass{cDSS2e} 
 
\usepackage{amsfonts} 
\usepackage{amsmath} 
\usepackage{xspace} 
\usepackage{color,graphicx} 
\usepackage{epsf} 
\usepackage{epsfig} 
\usepackage{psfrag} 
\usepackage{subfigure} 
 
\usepackage{cite}

\newcommand{\etal}{{\it et al.}\@\xspace} 

\newcommand{\half}{\frac{1}{2}}
\newcommand{\quarter}{\frac{1}{4}}
\newcommand{\fifth}{\frac{1}{5}}

\begin{document}

\doi{} 
 \issn{} 
\issnp{} \jvol{} \jnum{} \jyear{} \jmonth{} 
\markboth{}{} 

\title{Chaos in the Takens--Bogdanov bifurcation with $O(2)$ symmetry}

\author{A. M. Rucklidge\thanks{a.m.rucklidge@leeds.ac.uk}, 
Department of Applied Mathematics, University of Leeds, 
\break       Leeds LS2 9JT, UK 
\break E.~Knobloch,\thanks{knobloch@berkeley.edu}
Department of Physics, University of California, 
\break
Berkeley CA 94720, USA} 

\received{\today -- DOI:} 
\maketitle 

\begin{abstract}
The Takens--Bogdanov bifurcation is a codimension two bifurcation that provides a key to 
the presence of complex dynamics in many systems of physical interest. When the system 
is translation-invariant in one spatial dimension with no left-right preference the 
imposition of periodic boundary conditions leads to the Takens--Bogdanov bifurcation 
with $O(2)$ symmetry. This bifurcation, analyzed by G. Dangelmayr and E. Knobloch, Phil. 
Trans. R. Soc. London A {\bf 322}, 243 (1987), describes the interaction between steady 
states and traveling and standing waves in the nonlinear regime and predicts the presence
of modulated traveling waves as well. The analysis reveals the presence of several global 
bifurcations near which the averaging method (used in the original analysis) fails. We show here, using a combination of 
numerical continuation and the construction of appropriate return maps, that near the 
global bifurcation that terminates the branch of modulated traveling waves, the normal form for the Takens--Bogdanov bifurcation admits 
cascades of period-doubling bifurcations as well as chaotic dynamics of Shil'nikov type. 
Thus chaos is present arbitrarily close to the codimension two point.

\end{abstract}

%\begin{keyword}
%funny number/letter codes needed\newline
%Chaos, Takens--Bogdanov bifurcation, $O(2)$ symmetry.
%\end{keyword}

% 34K18 Bifurcation theory
% 34K23 Complex (chaotic) behavior of solutions            
% 34C37 Homoclinic and heteroclinic solutions           
                               
% cover letter                                   
                                                 
% Dear Editor,      
%                       
% Please consider this manuscript for publication
% in Dynamical Systems. In this paper, we describe chaotic dynamics 
% that occurs (surprisingly) in the normal form for the Takens-Bogdanov 
% bifurcation with O(2) symmetry. We find parameter values where a pitchfork
% and global bifurcation occur simultaneously and reduce the dynamics to a 
% two-dimensional map near this codimension-two point.                         
%                                                                              
% We should say that the work for this paper began in 1993, but it was only 
% relatively recently that we were able to find an appropriate limit in which   
% we could demonstrate a period-doubling (the first of a cascade leading to     
% chaos) in the map. 
% 
% Yours sincerely,
% 
% Alastair Rucklidge & Edgar Knobloch

\section{Introduction}
\label{secintro} 

When a layer of fluid is heated from below, steady convection sets in
once the temperature difference across the layer exceeds a critical value,
destabilizing the fluid by buoyancy effects. If sufficiently strong competing 
stabilizing effects are included the initial instability can be oscillatory~\cite{Chandrasekhar1961}, 
and the associated dynamics become much more interesting. This behavior is
organized by the Takens--Bogdanov (TB) bifurcation that occurs when the 
primary bifurcation changes from a pitchfork bifurcation (leading to 
steady convection) to a Hopf bifurcation (leading to oscillatory convection 
with the same wavenumber). This codimension-two bifurcation arises in a wide variety of convection 
problems, including thermosolutal convection, magnetoconvection, binary fluid convection and rotating
convection~\cite{Knobloch1981b,DaCosta1981,Guckenheimer1983a,Knobloch1986e,Rucklidge1992,Rucklidge1993a}, 
as well as other problems of physical interest such as Langmuir 
circulation~\cite{Cox1992}, coupled oscillators~\cite{Crawford1994b}, and the
formation of spiral waves in chemical reactions~\cite{Golubitsky1997}.

The linearized dynamics at the Takens--Bogdanov bifurcation are characterized
by the matrix $\Bigl[{0\atop0}\,{1\atop0}\Bigr]$, but the exact form of the
normal form equations that describe the nonlinear dynamics near the
codimension-two bifurcation point depends critically on the symmetry of the
problem. Even without any symmetry, the existence of a global (homoclinic)
bifurcation may be deduced, and the case of simple reflection ($Z_2$) symmetry
is also well understood~\cite{Knobloch1981b,Guckenheimer1983b}. Neither of
these cases exhibits chaos since the normal forms are two-dimensional. However,
even relatively simple extensions (for example, to $D_3$ and $D_4$
symmetry~\cite{Matthies1999,Rucklidge2001}) do exhibit chaotic dynamics within
the corresponding normal form although the details remain to be fully
understood.

We are interested here in the case of circular $O(2)$~symmetry, which is
relevant to two-dimensional convection with periodic boundary conditions in the
horizontal. The analysis of Dangelmayr and Knobloch \cite{Dangelmayr1987b} 
(hereafter DK) of the fourth-order normal form for this problem revealed a wide 
variety of competing states, including standing waves (SW), traveling waves (TW), 
modulated waves (MW) and steady-state convection (SS), and catalogued possible transitions
between these states and the trivial (zero) solution. However, DK found no chaotic 
dynamics near the codimension-two point, which is at first glance surprising given 
that the ordinary differential equations (ODEs) describing the dynamics are of
fourth order. However, one phase variable decouples (reducing the order to
three), and DK used an averaging procedure to reduce the order from three to
two, which excludes the possibility of chaotic dynamics. This procedure gives the
correct description of the dynamics close to the codimension-two bifurcation
point, but it makes assumptions about the time scales in the problem that are
not valid near global bifurcations. This does not cause problems in the case of
global bifurcations of SW, where the dynamics remain confined to a 
two-dimensional reflection-invariant subspace, but one of the situations they
discussed, in which a MW is created in a Hopf bifurcation from a TW, and is
then destroyed when it collides with the same TW in a global bifurcation, is
topologically impossible in second-order problems. 

\begin{figure}
 \begin{center}
 \mbox{\psfig{file=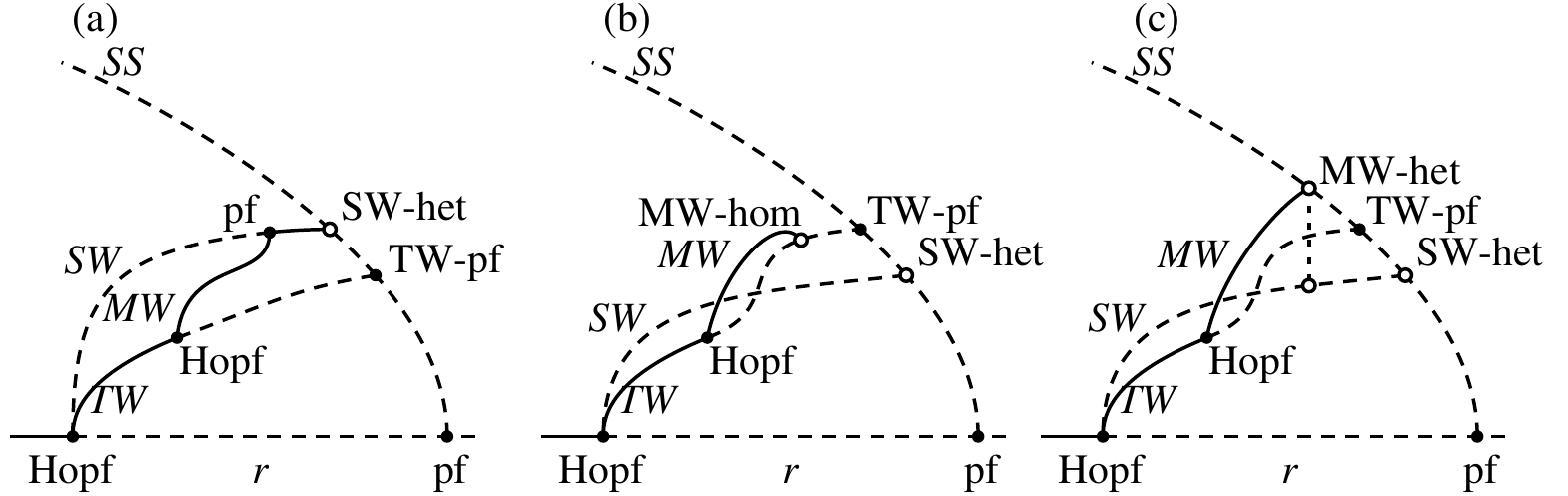,width=\hsize}}
 \end{center}
 \caption{Three scenarios for the ending of the MW branch:
 (a)~in a pitchfork bifurcation with the SW branch;
 (b)~in a global bifurcation with the TW branch;
 (c)~in a global bifurcation with the SW and SS branches. Diagrams 
 (a) and (b) are characteristic of parameter regions labeled 
 III$^-$ and II$^-$ by~DK \cite[figure~5]{Dangelmayr1987b},
 although (b) turns out not to be correct;
 (c) is as in \cite{Knobloch1990b}.}
 \label{figthreeendings}
 \end{figure}

In this paper, we examine in some detail the ways in which the branch of MW can
terminate. The two possibilities described by DK are that it ends in a pitchfork
bifurcation on the SW branch (Figure~\ref{figthreeendings}a) or in a global
bifurcation when it collides with TW (Figure~\ref{figthreeendings}b). We show
that this global bifurcation is associated with chaotic dynamics in the normal
form (and hence in physical problems described by the normal form) arbitrarily
close to the codimension-two point. We also find that farther from the
codimension-two point, the branch of MW can end in a global bifurcation
involving the SS and SW solutions (Figure~\ref{figthreeendings}c); this
scenario has been found in~\cite{Knobloch1990b,Knobloch1991,Cox1992}, but it
was not previously known to occur in the normal form. However, this global
bifurcation cannot be brought arbitrarily close to the codimension-two point,
and so is not strictly within the realm of validity of the normal form.

This paper is organized as follows. In section~\ref{secEquations} we summarize 
the normal form equations and their basic properties; in section~\ref{secResults} 
we describe our results obtained by a combination
of numerical continuation using AUTO and direct integration of the ODEs, and 
in section~\ref{secCrossing} we construct a two-dimensional map in order to analyse the dynamics near the coincidence of 
the pitchfork bifurcation from SS to TW and the heteroclinic bifurcation in 
which the SW collide with the SS equilibria. In section~\ref{secDiscussion} 
we discuss the significance of our findings.

\section{Equations}
\label{secEquations}

In the motivating physical problems, 
the state that bifurcates from the trivial solution is characterized by a
complex amplitude $v(t)$ multiplying the marginal eigenfunction with horizontal
wavenumber~$k$. In a periodic domain of width $\lambda=2\pi/k$ invariance of
the problem under translations ($x\rightarrow x+d$) and left--right reflection
($x\rightarrow \lambda-x$) induces the following action of $O(2)$ on the
amplitude~$v$:
 \begin{equation}\label{eqsymmetries}
 v\rightarrow {\rm e}^{{\rm i}kd}v,\qquad v\rightarrow{\bar v}.
 \end{equation}
The normal form of the Takens--Bogdanov bifurcation with $O(2)$ symmetry,
truncated at cubic order, is:
 \begin{equation}\label{eqnf}
 {\ddot v}=\mu v + \nu {\dot v} +
           A|v|^2v + B|{\dot v}|^2v + C(v{\dot{\bar v}} + {\bar v}{\dot v})v +
           D|v|^2{\dot v},
 \end{equation}
where $\mu$ and $\nu$ are unfolding parameters, and $A$, $B$, $C$ and $D$ are 
real constants~\cite{Dangelmayr1987b}. We suppose that we are close to the codimen\-sion-two 
TB~point $(\mu,\nu)=(0,0)$ and so explicitly scale $\mu$ and
$\nu$ by $\epsilon^2$, $v$ by $\epsilon$ and time by $\epsilon^{-1}$, where
$\epsilon\ll1$. The resulting scaled equation is
 \begin{equation}\label{eqnfscaled}
 {\ddot v}=\mu v + A|v|^2v + \epsilon\left(\nu {\dot v} +
           C(v{\dot{\bar v}} + {\bar v}{\dot v})v +
           D|v|^2{\dot v}\right) + {\mathcal O}(\epsilon^2).
 \end{equation}
In this paper only the dynamics that persist in the limit
$\epsilon\rightarrow0$ will be of interest. Equation (\ref{eqnfscaled}) shows that 
in this limit, the coefficient~$B$ drops out and does not affect the states we study or their stability.
In most of the analysis below, we find it convenient to approach this limit by
setting $\nu=\pm1$ and allowing $\epsilon$ to be small; effectively, we use
$(\mu,\epsilon)$ as unfolding parameters, and the original parameters in this
case are $(\epsilon^2\mu,\pm\epsilon^2)$.

In addition, we find it more convenient to work with the equations in the
form of a third-order system of ODEs for the variables $(r,s,L)$,
 \begin{equation}\label{eqnfrL}
 {\dot r}=s,
 \qquad
 {\dot s}=\mu r + Ar^3 + \epsilon\left(\nu + Mr^2\right)s
            + \frac{L^2}{r^3},
 \qquad
 {\dot L}= \epsilon L \left(\nu + Dr^2\right),
 \end{equation}
 where $M=2C+D$ and we have written $v=r{\rm e}^{{\rm i}\phi}$ with $L\equiv r^2{\dot\phi}$~\cite{Dangelmayr1987b}.
 This form of the problem is a consequence of rotational invariance of Eq.~(\ref{eqnfscaled}).

\begingroup
%\squeezetable
\begin{table}
\begin{tabular}{l|l}
\strut
Solution (Abbreviation) & Properties \\
\hline
\strut
Steady state (SS) & $(r,s,L)=(*,0,0)$ \\
\hline
\strut
Traveling wave (TW) & $(r,s,L)=(*,0,*)$ \\
\hline
\strut
Standing wave (SW) & $(r,s,L)=(**,**,0)$ \\
\hline
\strut
Modulated wave (MW) & $(r,s,L)=(**,**,**)$ \\
\hline
 \end{tabular}
 \caption{\label{tab:solutions}Properties of the four basic solutions 
of Eq.~(\ref{eqnfrL}), where $*$ indicates a constant non-zero value, while $**$ 
indicates a time-dependent non-zero value.}
 \end{table}
 \endgroup

The four known types of solution (see table~\ref{tab:solutions})
are steady-states (SS), for which $s=L=0$ 
and $r=\sqrt{-\mu/A}$, traveling waves (TW), for which $s={\dot L}=0$,
$r=\sqrt{-\nu/D}$ and $L=\pm\frac{\nu}{D}\sqrt{-(\mu-(A/D)\nu)}$, standing waves
(SW), which are periodic orbits with $L=0$, and modulated waves (MW), which are
periodic orbits with $L\neq0$.

In the limit of small~$\epsilon$, DK derived averaged equations 
for $E\equiv \half s^2 + \half\frac{L^2}{r^2} - \half\mu r^2 - \quarter Ar^4$
and~$L$,
 \begin{equation}\label{eqnEL}
 {\dot E} = \epsilon f(E,L^2) + \mathcal{O}(\epsilon^2)
 \qquad\text{and}\qquad
 {\dot L} = \epsilon L g(E,L^2) + \mathcal{O}(\epsilon^2),
 \end{equation}
where the functions $f$ and $g$ were expressed as elliptic integrals. The SS,
SW, TW and MW solutions were found as roots of $f=Lg=0$. Truncated at
$\mathcal{O}(\epsilon)$, these second-order ODEs cannot have chaotic solutions.
The derivation of~(\ref{eqnEL}) relied on a separation of time-scales between
the period of closed orbits in~(\ref{eqnfrL}) and the time-scale over which
these orbits evolve. This separation does not hold near the homoclinic
bifurcations of MW since the period of the MW diverges to infinity; it is
near the homoclinic bifurcations of MW that we find chaotic solutions.

We recall briefly some of the relevant results of \hbox{DK}. The trivial
solution $v=0$ is stable when $\mu$ and $\nu$ are both negative; it loses
stability in a pitchfork bifurcation (creating a circle of SS) when $\mu=0$ and
in a Hopf bifurcation (creating SW and TW) when $\nu=0$ with $\mu<0$. The Hopf
bifurcation from TW to MW (TW-Hopf) occurs when
 \begin{equation}\label{eqnHopfTWMW}
 \mu=\frac{3M-5D}{2M-4D}\frac{A}{D}\nu,\qquad\hbox{with }\mu<\frac{A}{D}\nu,
 \end{equation}
while the pitchfork bifurcation from SS to TW (SS-pf) occurs when
 \begin{equation}\label{eqnpfSSTW}
 \mu=\frac{A}{D}\nu,\qquad\hbox{with }A\mu<0.
 \end{equation}
The SW solution is destroyed in a heteroclinic bifurcation (SW-het) when it
collides with the SS equilibria at
 \begin{equation}\label{eqnhetSWSS}
 \mu=\frac{5A}{M}\nu+{\mathcal O}(\epsilon),\qquad\hbox{with }A>0,\,\mu<0,
 \end{equation}
while the MW solution is destroyed in a homoclinic bifurcation (MW-hom) when it
collides with the TW branch at
 \begin{equation}\label{eqnhomMWTW}
 \mu=\frac{3M-5D}{2M}\frac{A}{D}\nu+{\mathcal O}(\epsilon),
  \qquad\hbox{with }A>0,\,\mu<0\hbox{ and }0<\frac{D}{M}<\fifth.
 \end{equation}
Finally, there is the possibility of a pitchfork bifurcation from SW to MW 
(SW-pf). The expression for the location of this bifurcation involves computing
elliptic integrals, but in the special case $\frac{D}{M}=\fifth$, it occurs along the
half-line
 \begin{equation}\label{eqnpfSWMW}
 \mu=\frac{A}{D}\nu+{\mathcal O}(\epsilon),
\qquad\hbox{with }A>0,\,\mu<0 \hbox{ and }\frac{D}{M}=\fifth.
 \end{equation}
Thus in the special case $\frac{D}{M}=\fifth$ the last four of these bifurcations
(SS-pf, SW-het, MW-hom and SW-pf) occur along lines that are tangent to
$\mu=A\nu/D$ as $\epsilon\rightarrow0$. This four-fold degeneracy is lifted
when $\epsilon$ is finite, i.e., at any finite distance from the TB~point, or
by changing the value of $D/M$. To understand the transition between regions
II$^-$ and III$^-$ in DK~\cite[figure~5]{Dangelmayr1987b} we need, therefore,
to go beyond the analysis of~\hbox{DK} and understand how the degeneracy is
lifted for small but finite values of~$\epsilon$, i.e., we need to determine
the role of the higher order terms in eqs.\
(\ref{eqnpfSSTW})--(\ref{eqnpfSWMW}), and to investigate which other
bifurcations or solutions might occur.

 %While these lines are tangent to each other at the codimen\-sion-two point,
 %this is no longer so when the parameters are moved away from this point since
 %the coincidence of four bifurcations goes beyond the degeneracy one would
 %expect from choosing a single special value of~$D/M$. The higher-order terms
 %dropped from the above expressions destroy the degeneracy, and at any finite
 %distance from the TB~point, the four bifurcations no longer coincide.

Before discussing the new results, we consider the issue of the stability of
the MW at the homoclinic bifurcation when they collide with~\hbox{TW}.
DK~discussed the two Floquet multipliers of the MW derived from the
second-order averaged equations~(\ref{eqnEL}), and concluded that in the case
$A>0$, $D<0$, $M<0$ and $\fifth<\frac{D}{M}<\half$, stable MW are created in a
supercritical Hopf bifurcation from TW, with the branch of stable MW terminating
in a pitchfork bifurcation on SW (figure~\ref{figthreeendings}a). In the case
$0<\frac{D}{M}<\fifth$, there is still a branch of stable MW created in a supercritical
Hopf bifurcation from TW, but the MW are destroyed in a homoclinic
bifurcation on the TW branch (figure~\ref{figthreeendings}b). Within the
second-order averaged equations, the MW are supposed to be stable at the
homoclinic bifurcation since the negative eigenvalue of TW is greater in
magnitude than the positive. However, in the third-order
equations~(\ref{eqnfrL}), between the Hopf and homoclinic bifurcations, the TW
have one negative real eigenvalue and two eigenvalues with positive real part
(since they started stable and then underwent a Hopf bifurcation), and at the
point of the MW-hom bifurcation, one can show that the two positive eigenvalues
are real. Generically, periodic orbits that approach a homoclinic bifurcation 
involving an equilibrium point with one negative and two distinct positive real 
eigenvalues are not stable~\cite{Kuznetsov1998}, and so the
MW cannot be stable at the homoclinic bifurcation. The change from stable MW
(at the Hopf bifurcation) to unstable MW (at the homoclinic bifurcation) was
therefore not captured by the analysis of~\hbox{DK}.

These considerations reveal that there are two issues to be resolved in
understanding of the Takens--Bogdanov normal form with $O(2)$ symmetry: how do
the four coincident bifurcations split apart when higher-order terms are taken
into consideration, and how do the MW become unstable before they collide with
the TW branch. To explore these issues, we focus on the interesting case $A=1$,
$D=-1$ and $M\approx -5$. There is a third consideration, which is how the
homoclinic bifurcation of MW changes into a heteroclinic bifurcation involving
SS and SW further away from the codimen\-sion-two point, a situation we
consider briefly at the end of section~\ref{secResults}.

 \begin{figure}
 \begin{center}
 \mbox{\psfig{file=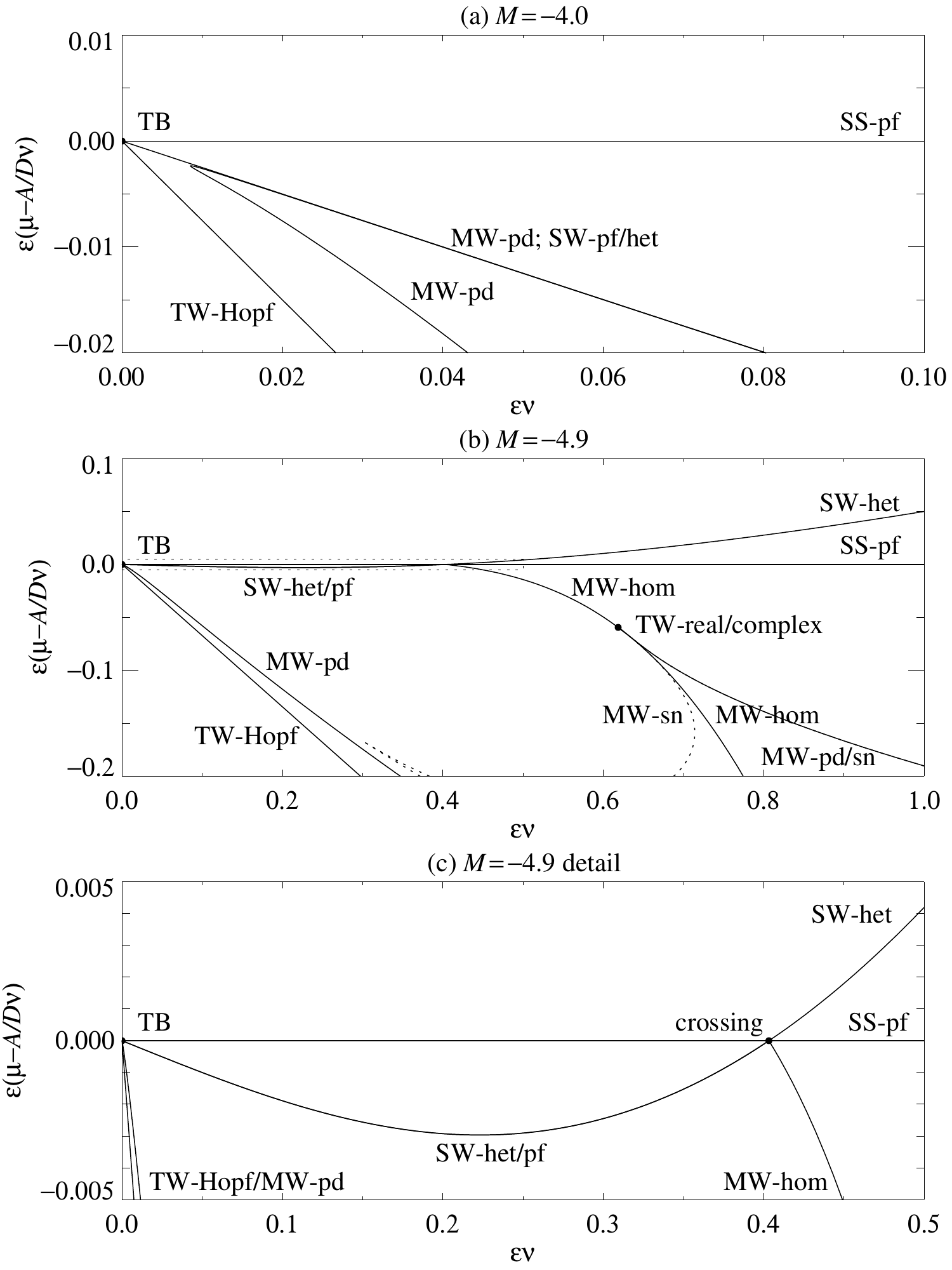,width=\hsize}}
 \end{center}
 \caption{Computed unfolding diagrams for Eqs.~(\ref{eqnfrL}), with $A=1$ and $D=-1$, 
and
 (a)~$M=-4.0$, 
 (b)~$M=-4.9$, 
 (c)~$M=-4.9$ (detail of the dotted box region in~b), 
 We plot $\epsilon\nu$ on the horizontal axis and
$\epsilon(\mu-\frac{A}{D}\nu)$ on the vertical axis, so that the line of SS-pf
is horizontal. Some of the lines are too close together (for example, SW-pf/SW-het, and MW-pd/MW-sn) to be resolved in this
diagram; a schematic drawing of the unfolding diagram is therefore shown in
figure~\ref{figunfNearMminusfive}. The crossing of the SS-pf and SW-het
bifurcations is clearly seen in~(b), and in more detail in~(c).
The MW-hom line emerges from the crossing point.}
 \label{figcrossing1}
 \end{figure}

 \begin{figure}
 \begin{center}
 \mbox{\psfig{file=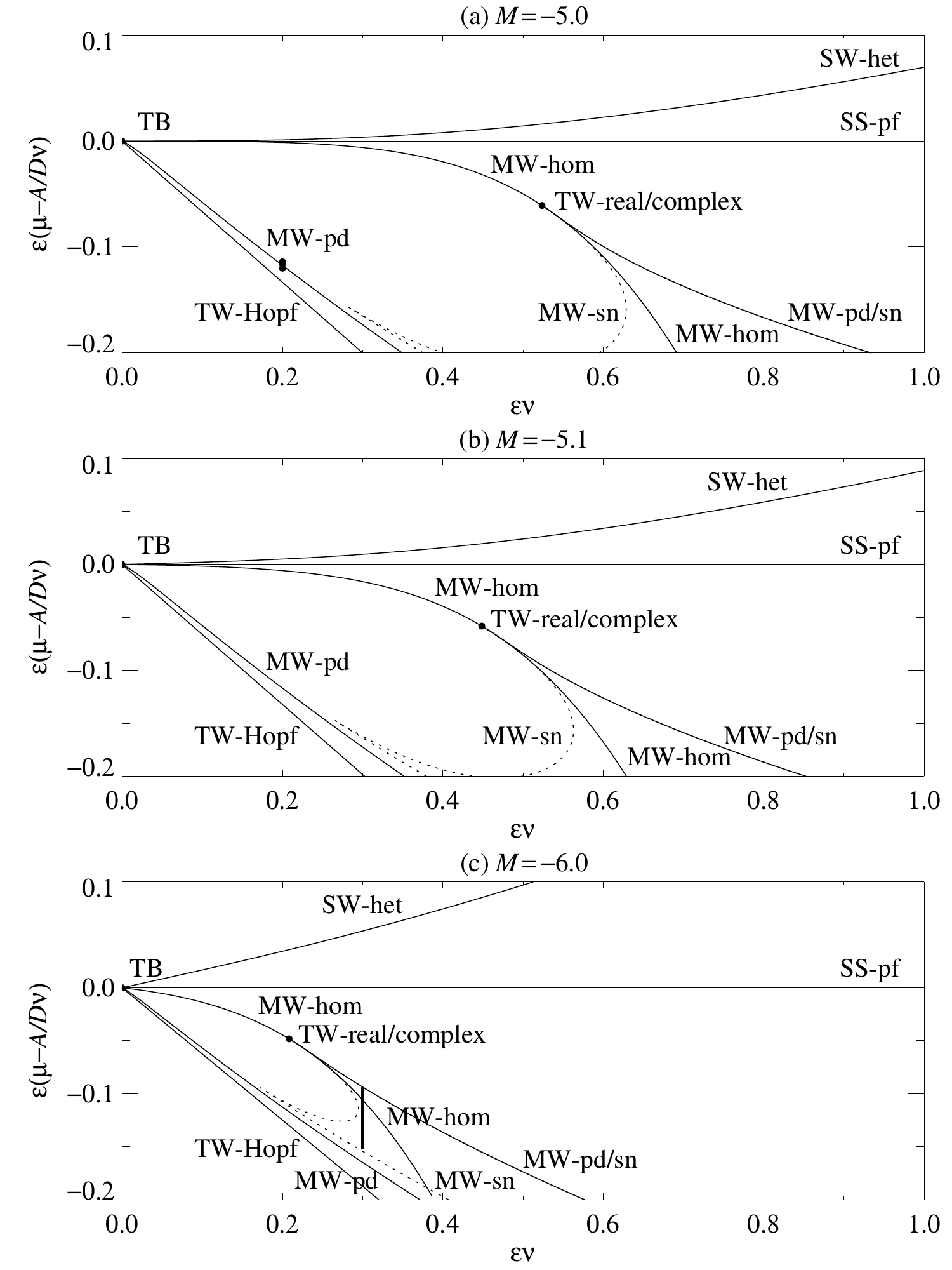,width=\hsize}}
 \end{center}
 \caption{Continuation of figure~\ref{figcrossing1}: 
computed unfolding diagrams for Eqs.~(\ref{eqnfrL}), with $A=1$ and $D=-1$, 
and
 (a)~$M=-5.0$, 
 (b)~$M=-5.1$, 
 (c)~$M=-6.0$.
In~(a), with $M=-5.0$, the lines of SS-pf, SW-het and MW-hom are
tangent at the TB~point, and the line of SW-pf has shrunk to zero.
On the MW-hom line, where MW are homoclinic to TW, the leading eigenvalues of 
TW change from being real to complex, and secondary lines of MW-sn and MW-pd 
(demarcating the edges of the MW period-doubling cascade) emerge.
In (a), the dots at $\epsilon\nu=0.2$, below the `MW-pd' label, indicate the 
parameter values in figure~\ref{figexamples1}. In (c), the vertical bar at
$\epsilon\nu=0.3$ indicates the range between the leftmost and rightmost
saddle-node bifurcations in figure~\ref{figexamplesmwhet}(b).}
 \label{figcrossing2}
 \end{figure}

%  \begin{figure}
%  \begin{center}
%  \mbox{\psfig{file=pd_approach,width=\hsize}}
%  \end{center}
%  \caption{Numerically determined location of the period-doubling bifurcation 
% from~\hbox{MW} (solid line: MW-pd).
% The graphs show the value of~$\mu$ (which has been scaled 
% with~$\epsilon^2$) at which the MW-pd bifurcation 
% occurs as~$\epsilon$ is varied: the value of~$\mu$ as $\epsilon\rightarrow0$
% gives the slope of the bifurcation line in the unscaled unfolding diagrams. The 
% dash-dot lines are the locations of the SW-het bifurcation, and the 
% dashed lines show an extrapolation of the MW-pd curve to 
% $\epsilon\rightarrow0$.
%  (a)~$M=-4.9$, with the MW-pd bifurcation for $M=-4.0$ included as a dotted 
% line.
%  (b--d)~$M=-5.0$, $-5.1$ and~$-6.0$.}
%  \label{figpdapproach}
%  \end{figure}

 \begin{figure}
 \begin{center}
 \mbox{\psfig{file=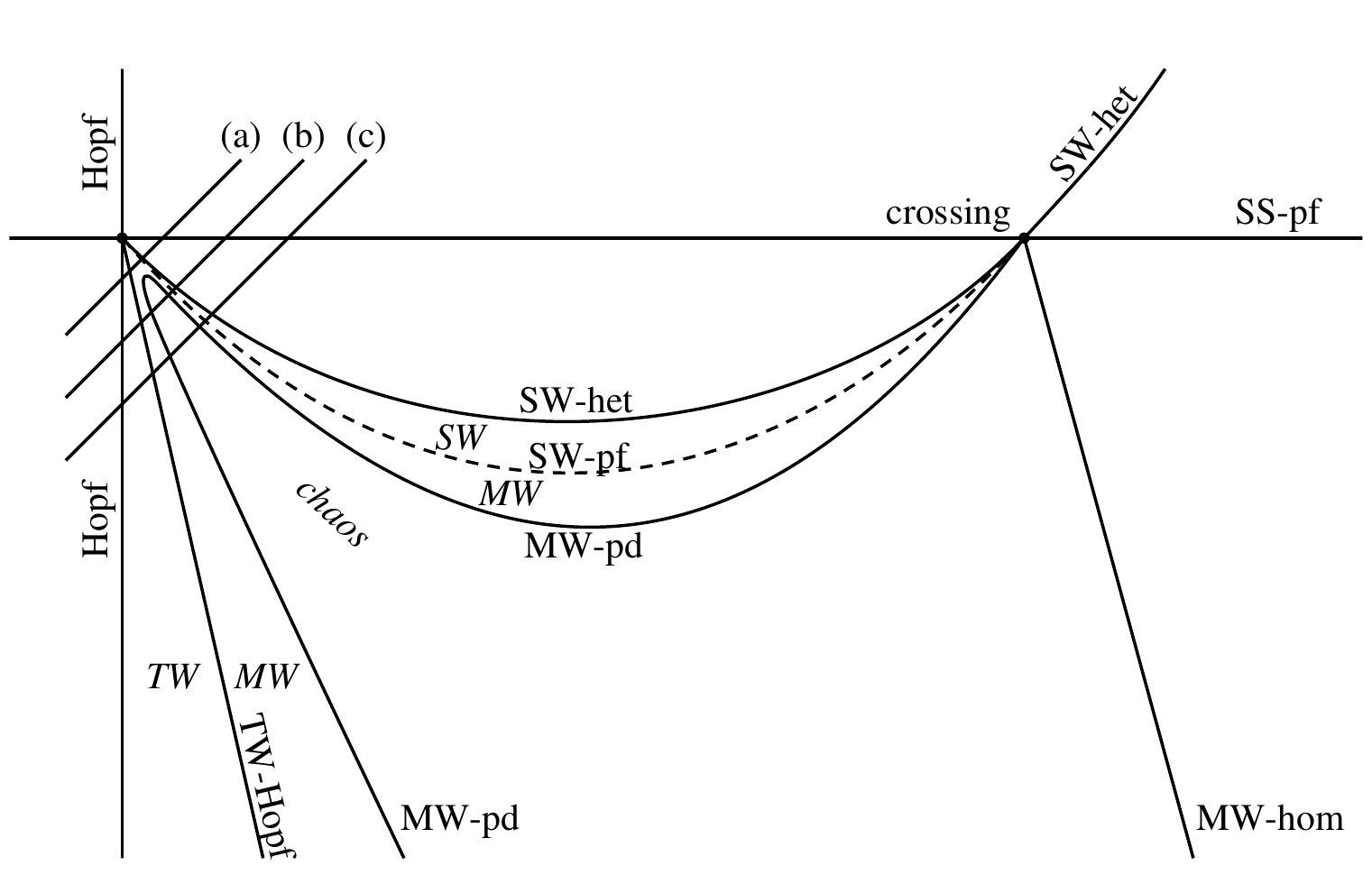,width=\hsize}}
 \end{center}
 \caption{Schematic drawing of the unfolding diagram, for $M$~just above~$-5$,
and including the crossing of the SS-pf and SW-het bifurcations. The loop of
MW-pd bifurcations comes close to the TB~point, but does not connect to it,
instead turning away and apparently ending at the crossing
point, where the line of MW-hom bifurcations begin. As $M$~decreases
below~$-5$, the crossing point moves to the left and vanishes when it collides
with the TB~point, allowing the MW-pd and MW-hom bifurcation lines to
connect to the TB~point (see figure~\ref{figcrossing2}b,c).
Regions of stable solutions (TW, MW, chaos and SW) are indicated in italics.}
 \label{figunfNearMminusfive}
 \end{figure}

 \begin{figure}
 \begin{center}
 \mbox{\psfig{file=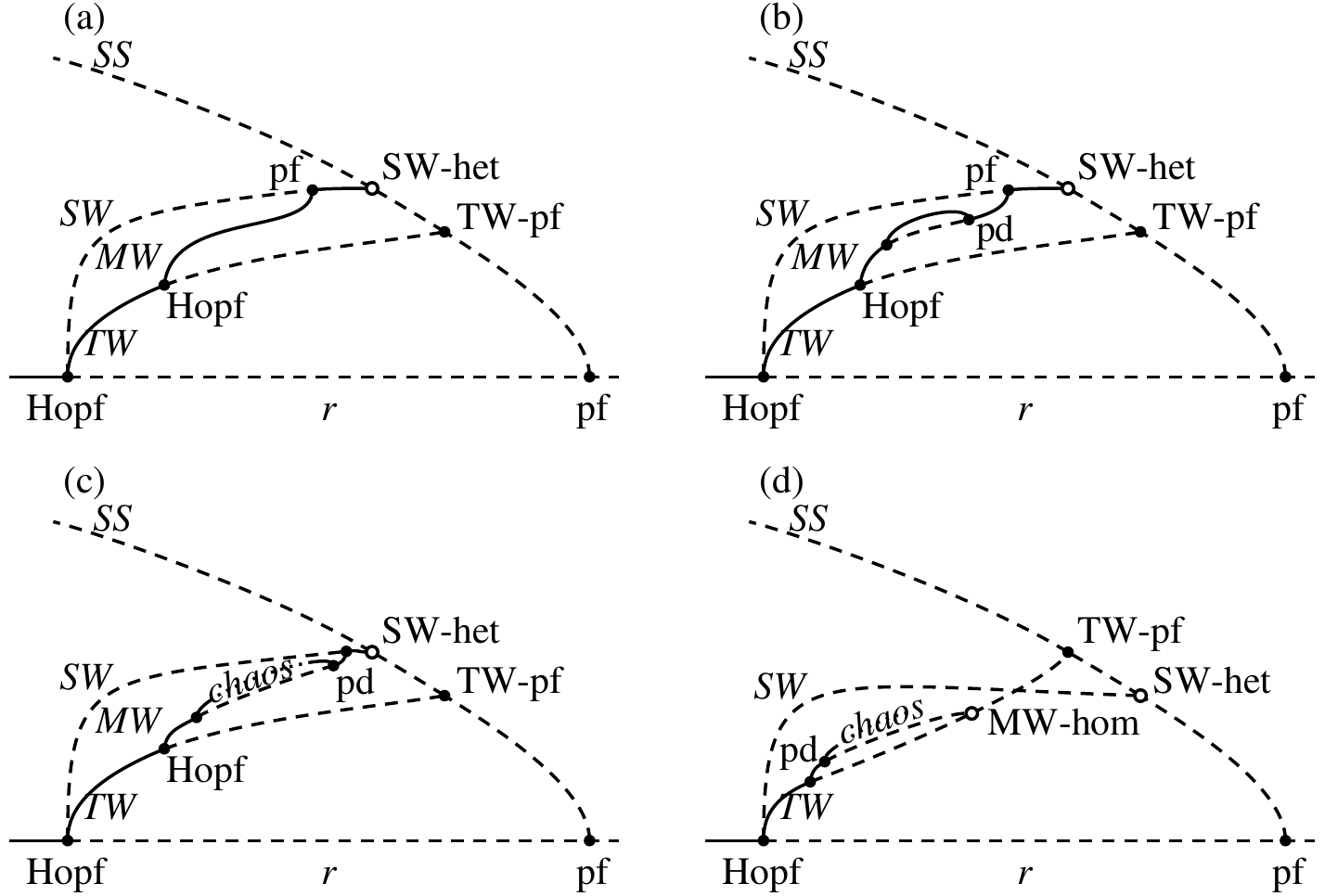,width=\hsize}}
 \end{center}
 \caption{Schematic bifurcation diagrams taken along the cuts indicated in
figure~\ref{figunfNearMminusfive}. (a)~and~(b): with $M=-4.0$, taken close to the
TB~point (see figure~\ref{figthreeendings}a), and slightly further
away, showing how an interval of MW-period doubling comes in. (c):~with
$M=-4.9$, there is a period-doubling cascade leading to chaotic MW
trajectories, but MW must be restabilized before the branch ends in the SW-pf
bifurcation; this occurs through a MW-pd bifurcation. (d):~with $M=-5.1$, the
SW-het and TW-pf bifurcations have changed order on the SS branch. The MW
branch ends in the MW-hom bifurcation when unstable MW collide with TW, but the
chaotic MW persist (contrast with figure~\ref{figthreeendings}b).}
 \label{figbifNearMminusfive}
 \end{figure}                 

 \begin{figure}
 \begin{center}
 \mbox{\psfig{file=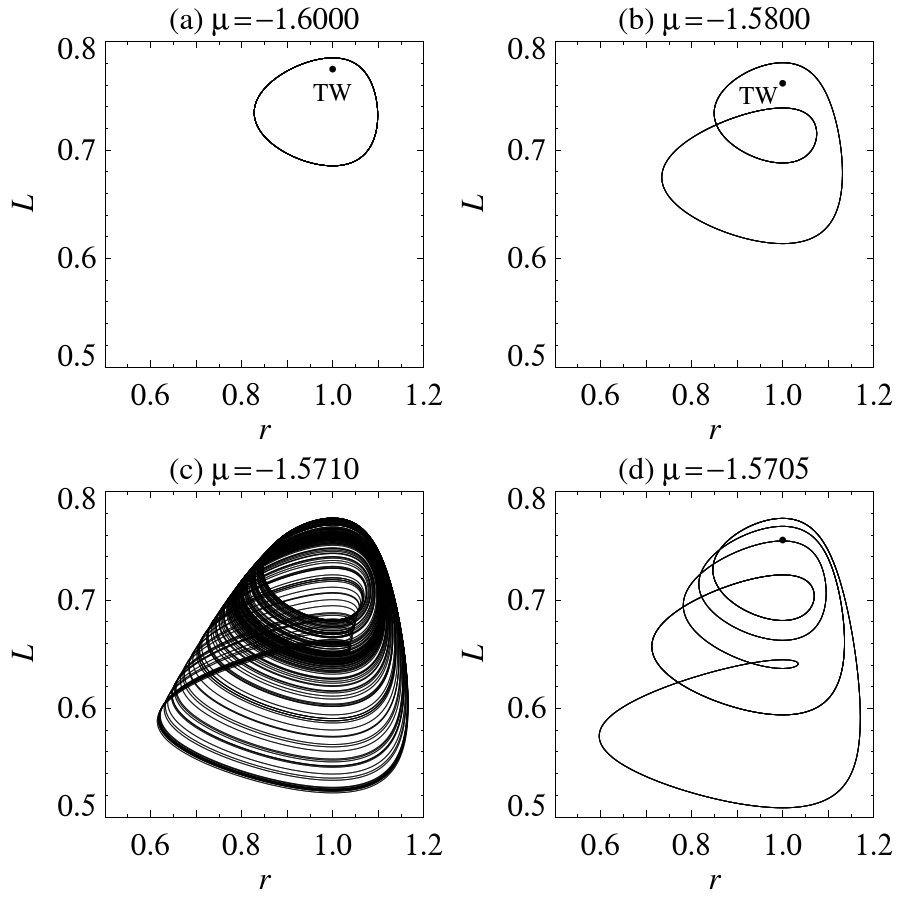,width=\hsize}}
 \end{center}
 \caption{Examples of stable solutions of Eqs.~(\ref{eqnfrL}) 
in the $(r,L)$ plane, with 
$A=1$, $M=-5$, $D=-1$, $\nu=1$, $\epsilon=0.2$ and 
(a)~$\mu=-1.6000$ (stable MW),
(b)~$\mu=-1.5800$ (period-doubled MW),
(c)~$\mu=-1.5710$ (chaotic MW),
(d)~$\mu=-1.5705$ (MW in a period 5 window).
This range of parameter values is indicated in figure~\ref{figcrossing2}(a).
The TW equilibrium point is indicated.}
 \label{figexamples1}
 \end{figure}                 

 \begin{figure}
 \begin{center}
 \mbox{\psfig{file=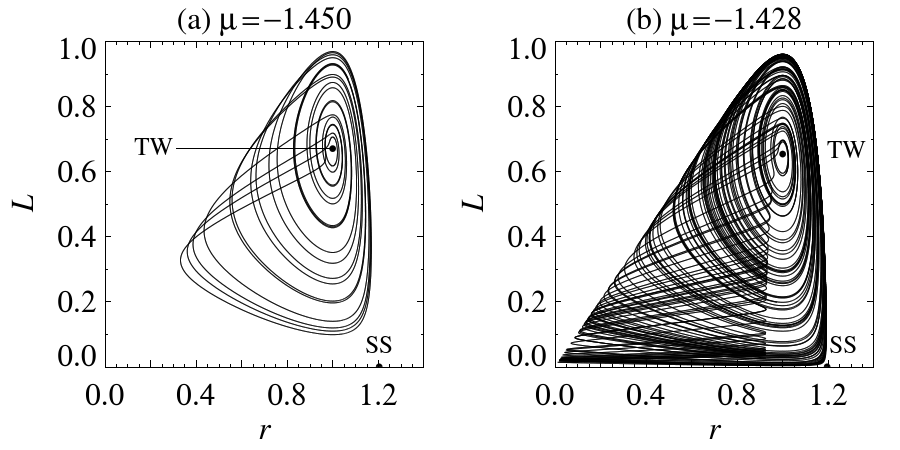,width=\hsize}}
 \end{center}
 \caption{Examples of stable solutions of Eqs.~(\ref{eqnfrL}) in the 
$(r,L)$ plane, with $A=1$, $M=-5$, $D=-1$, $\nu=1$, $\epsilon=1$ and 
(a)~$\mu=-1.450$ (chaotic MW near a MW-homoclinic bifurcation with TW),
(b)~$\mu=-1.428$ (chaotic MW near a MW-heteroclinic bifurcation with TW, 
SS and SW);
these parameter values are outside the range of figure~\ref{figcrossing2}(a).
The TW and SS equilibrium points are indicated; SW lie in the plane $L=0$,
with $0\leq{r}\leq0.9$.}
 \label{figexamples2}
 \end{figure}                 

 \begin{figure}
 \begin{center}
 \mbox{\psfig{file=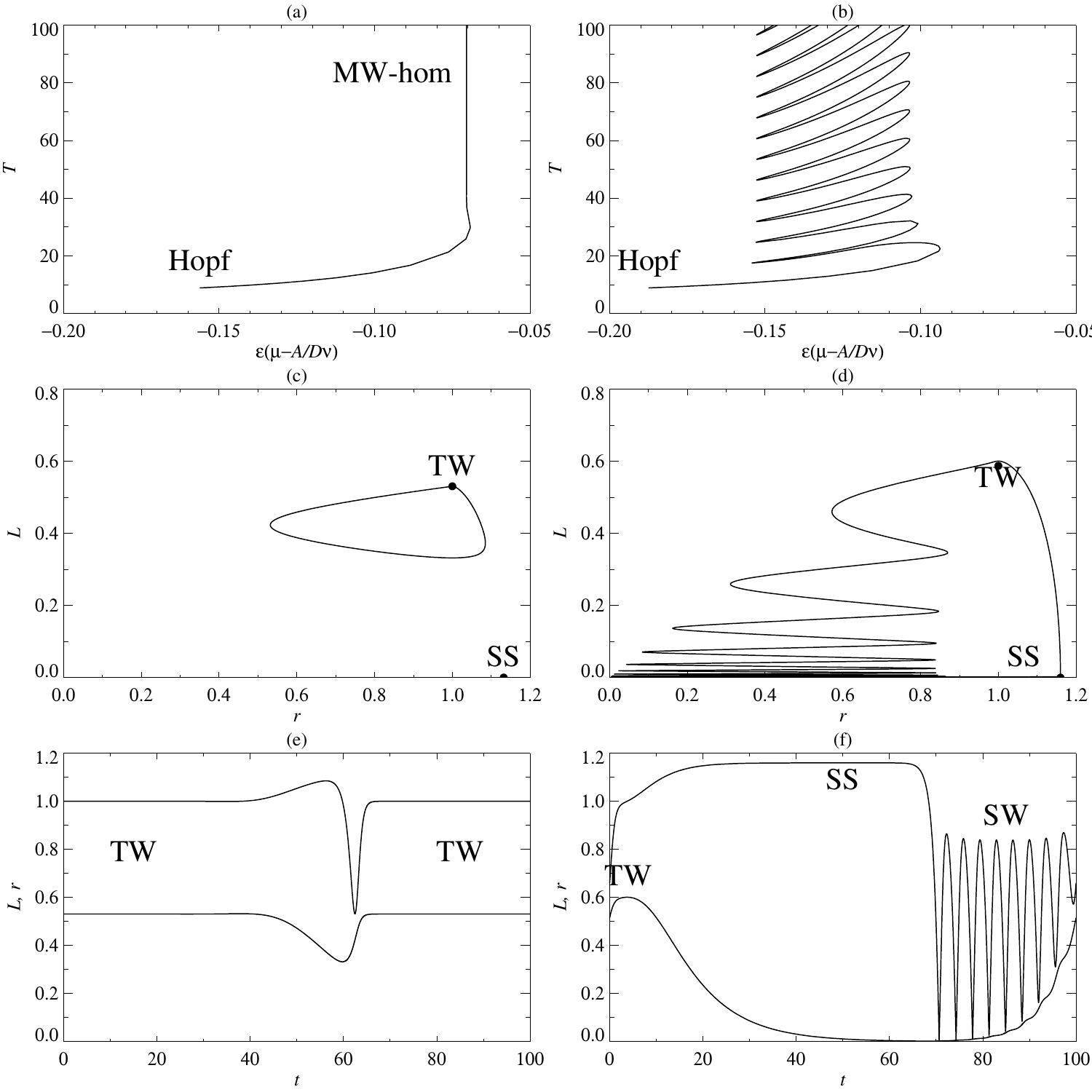,width=\hsize}}
 \end{center}
 \caption{Examples of MW solutions of Eqs.~(\ref{eqnfrL}) with $M=-6$ 
and the other parameters as in figure~\ref{figcrossing2}(c) found using 
AUTO. Left column: $\epsilon=0.25$; right column: $\epsilon=0.30$.
With $\epsilon=0.25$, the MW are produced at a Hopf bifurcation and 
terminate in a homoclinic bifurcation involving the TW saddle-focus 
equilibrium point. Panel (a) shows the period $T$ of the MW as a 
function of $\epsilon(\mu-\frac{A}{D}\nu)$.  The MW solution close to 
the MW-hom bifurcation is shown in (c) (with period $T=100$ and 
$\mu=-1.2816295$), and a time series is shown in (e), with the top line 
indicating $r(t)$ and the bottom line the corresponding~$L(t)$.
With $\epsilon=0.30$, the MW branch still originates in a Hopf bifurcation 
but subsequently begins to snake back and forth as the MW winds up around 
the SW periodic orbit (panel (b)). The extent of this snaking, i.e., the
distance between the leftmost and rightmost saddle-node bifurcations, is 
indicated in figure~\ref{figcrossing2}(c); there are also period-doubling 
bifurcations. The MW solution, shown in (d,f) with period $T=100$ and 
$\mu=-1.3451861$, approaches the TW and SS equilibria and the SW periodic
orbit (cf. \cite{Knobloch1990b,Knobloch1991}).}
 \label{figexamplesmwhet}
 \end{figure}                 

\section{Numerical results near the Takens--Bogdanov point}
\label{secResults}

We have carried out a series of numerical experiments on the
ODEs~(\ref{eqnfrL}) using the AUTO2000~\cite{Doedel2001} suite of continuation
software to follow several of the bifurcations, including the SW-het, MW-hom
and SW-pf bifurcations, as well as the first period-doubling of MW (MW-pd) and
a saddle-node of MW (MW-sn). For these computations we fixed $\nu=\pm1$ 
($\nu=+1$ is the interesting case) and used $\mu$ and $\epsilon$ as our 
unfolding parameters; the results below are presented as functions of 
$\epsilon\nu$ and $\epsilon(\mu-\frac{A}{D}\nu)$, chosen so that the SS-pf 
bifurcation line is horizontal.

The results are shown in figures~\ref{figcrossing1} and~\ref{figcrossing2}. 
Schematic version of the unfolding and bifurcation diagrams are shown in
figures~\ref{figunfNearMminusfive} and~\ref{figbifNearMminusfive}, and examples
of numerical solutions of (\ref{eqnfrL}), showing stable MW, period-doubled MW
as well as chaotic solutions, are in figures~\ref{figexamples1}
and~\ref{figexamples2}. Finally, figure~\ref{figexamplesmwhet} shows how the
MW-hom bifurcation changes to a heteroclinic bifurcation, with snaking
behavior, further from the TB~point.

These calculations were difficult to perform (especially for small~$\epsilon$),
and we took great care in choosing the various parameters in AUTO that control
the accuracy of the results. One reason for the difficulty is that some of the
bifurcations are exceedingly close to each other; furthermore, close to the
Takens--Bogdanov point, the $-1$ Floquet multiplier at the period-doubling
bifurcation is nearly degenerate, suggesting that there is a possibility of a
torus (Neimark--Sacker) bifurcation from the MW, although we were not able to
detect this with confidence. As a consequence of having a second Floquet
multiplier close to the unit circle, the stable MW (and chaotic solutions) are
only just stable for small~$\epsilon$, and when following solutions using
time-stepping, very small changes in parameter value were required if
trajectories are not to fall off the stable branch and diverge to infinity.

With $M>-5$ ($M=-4.0$ and $-4.9$ in figure~\ref{figcrossing1}a--c, shown
schematically in figures~\ref{figunfNearMminusfive}
and~\ref{figbifNearMminusfive}), the MW branch exists between the TW-Hopf and
the SW-pf bifurcations. The second of these is almost immediately followed by
the SW-het bifurcation, in which the SW branch is destroyed when in collides
with the SS equilibria, consistent with the analysis of DK
(figure~\ref{figthreeendings}a). However, it is clear from
figure~\ref{figcrossing1}(b,c) that the curves of SW-het and SW-pf begin below
the line of SS-pf but bend upwards and there is a codimen\-sion-two crossing
point at which these bifurcations coincide. We analyse this crossing point 
in detail in section~\ref{secCrossing}, but a summary is shown in
figure~\ref{figunfNearMminusfive}: the SW-het bifurcation continues through this 
point, while the SW-pf line terminates there and a line of MW-hom bifurcations begins
there. We also found a line of MW-pd bifurcations that begins at the crossing
point; we present stable period-doubled and chaotic MW at $M=-5$ in
figure~\ref{figexamples1}, and present evidence in section~\ref{secCrossing}
for the origin of this line.

Figure~\ref{figbifNearMminusfive} shows
schematic bifurcation diagrams taken along the cuts indicated schematically in
figure~\ref{figunfNearMminusfive}. These cuts are notionally taken 
across the unfolding diagram at a small but non-zero distance from 
the TB~point, and show how the chaotic dynamics inevitably approaches the 
TB~point as $M$~is decreased through~$M=-5$.

Figure \ref{figbifNearMminusfive}(a) shows the DK scenario.
Further from the TB point, a bubble of period-doubled MW develops
(figure~\ref{figbifNearMminusfive}b); further still, there is a 
period-doubling cascade leading to chaotic MW, followed by an
inverse cascade back to MW (figure~\ref{figbifNearMminusfive}c).

The DK scenario is always correct for $M>-5$, provided that the cut is taken 
progressively closer to the TB~point as $M$ approaches~$-5$. But, as $M$ passes 
below~$-5$ (figure~\ref{figbifNearMminusfive}d), the SW-het emerges from the TB 
point above instead of below to SS-pf line, implying the crossing point has moved 
into the TB~point, the SW-het and TW-pf bifurcations have changed order, and the 
SW-pf and final MW-pd bifurcations have vanished and been replaced by the MW-hom 
bifurcation, i.e., the location of the termination of the MW on the \hbox{TW} 
branch. However, the initial MW-pd bifurcation survives as does the period-doubling 
cascade leading to chaotic~\hbox{MW}. Figure~\ref{figbifNearMminusfive}(d) should 
be contrasted with the DK scenario (figure~\ref{figthreeendings}b), 
where there is no chaos. This is because the averaged equations used in DK
cannot predict the presence of MW period-doubling. The reason for this
difficulty is fundamental: the averaging technique replaces a two-torus
in~(\ref{eqnf}) by an equilibrium point in~(\ref{eqnEL}), and this equilibrium
point is identical for all trajectories on this torus, whether periodic or
quasiperiodic.

The secondary codimension-two bifurcation point, at which the SW-het and TW-pf
bifurcation lines cross, therefore holds the key to resolving the first of the
two difficulties with the DK results. The evidence presented so far shows that
the crossing point moves to the TB~point as $M$ decreases through $-5$, and
vanishes at $M=-5$. For $M<-5$, all the bifurcation lines that connect to the
crossing point will then connect to the TB~point instead. In 
section~\ref{secCrossing}, we show that this includes a MW-pd bifurcation, 
allowing a consistent interpretation of the numerical results (although we have 
been unable to continue numerically the line of MW-pd bifurcations all the 
way to the crossing point.)

We conclude this section by showing some period-doubled and chaotic solutions 
of Eqs.~(\ref{eqnfrL}) as well as homoclinic and heteroclinic solutions. In 
figure~\ref{figexamples1}, we take $M=-5$ and $\epsilon=0.2$, and show the 
progression from (a)~MW to (b)~period-doubled MW to (c)~chaotic~MW. We also 
show in figure~\ref{figexamples1}(d) a period 5 MW in a periodic window. The 
range of parameter values is indicated in figure~\ref{figcrossing2}(a). 
Although these computations are for $M=-5$, we find similar solutions with 
$M$ just above and just below~$M=-5$, thereby providing evidence for the region 
labeled `chaos' in the schematic diagrams in figure~\ref{figunfNearMminusfive} 
and the initial progress from stable MW to chaos summarized in 
figure~\ref{figbifNearMminusfive}(c,d). The termination of the MW branch is too 
delicate for numerical computation with very small~$\epsilon$, although we have 
been able to find chaotic solutions for $\epsilon$ as small as~$0.05$.

In figure~\ref{figexamples2}, we take $M=-5$ and $\epsilon=1$, further from the
TB point, and show (a)~chaotic solutions close to a Shil'nikov homoclinic
bifurcation involving the TW saddle-focus, and (b)~chaotic solutions close to a
heteroclinic connection between TW, SS and SW. For the parameter values in~(a),
the TW equilibrium point has eigenvalues $0.07627\pm0.65395i$ and $-4.15255$,
with contraction being greater than expansion, consistent with the chaos being
stable.

To understand the behavior in figure~\ref{figexamples2}(b) in more detail we
show in figure~\ref{figexamplesmwhet}(a,c,e) the MW solutions corresponding to
$M=-6$ and $\epsilon=0.25$. The figure shows that in this case the MW branch
terminates by forming a homoclinic connection with the TW equilibrium point.
Figure~\ref{figexamplesmwhet}(b,d,f) shows that when $\epsilon=0.30$ the MW
terminate instead in a heteroclinic connection with the SS equilibrium point
and the SW periodic orbit. Specifically, figure~\ref{figexamplesmwhet}(d,f)
demonstrates that the SW in the plane $L=0$ are unstable with respect to TW
perturbations; since the TW are themselves unstable to MW, the trajectory
spirals away from $L=0$ towards MW but the modulation takes the trajectory back
towards SS and hence the $L=0$ invariant plane. In this plane the SS is
unstable and solutions are attracted to SW. The process then repeats. Evidently
the chaotic solution is associated with the resulting connection from SW to SW such as occurs in the
Shil'nikov--Hopf bifurcation \cite{Hirschberg1993}. This generic situation is
modified here by the fact that the SW lie in the $L=0$ invariant plane with
robust trajectories from SS to SW. The associated bifurcation diagram in
figure~\ref{figexamplesmwhet}(b) exhibits snaking associated with the addition
of more and more turns around the SW limit cycle. The period~$T$ of the
associated MW diverges within a finite interval of $\epsilon(\mu-A/D\nu)$
instead of a single value as in figure~\ref{figexamplesmwhet}(a), a consequence
of the structural stability of the intersection between the 2D unstable
manifold of SW and the 2D stable manifold of the SW within $L=0$.

The change from homoclinic to heteroclinic behavior shown in
figure~\ref{figexamplesmwhet} occurs beyond the region of validity of the
normal form (it does not connect to the TB point), and is present for $M$ above
and below~$M=-5$. However, the resulting phase portraits greatly resemble those
identified in higher-dimensional systems \cite{Knobloch1990b,Knobloch1991} and
even in partial differential equations (PDEs) \cite{Cox1992}, suggesting that
similar transitions occur in these systems and that these may in fact be
captured by the normal form (\ref{eqnfrL}) provided that averaging is eschewed.
For this purpose it would be of particular interest to investigate the
properties of the full third-order normal form (\ref{eqnf}). In contrast, the
SW chaos observed in the PDEs describing double-diffusive convection is not
captured by this normal form since it takes place in the $L=0$ invariant
subspace -- the description of this type of chaos requires that the leading
eigenvalues of SS at SW-het are complex, something that occurs only a finite
distance from the TB codimension-two point \cite{Knobloch1986a}. Note that this
picture provides a natural construction for a homoclinic connection of a
strange invariant set in $L=0$ to itself \cite{Knobloch1991}.

 \begin{figure}
 \begin{center}
 \mbox{\psfig{file=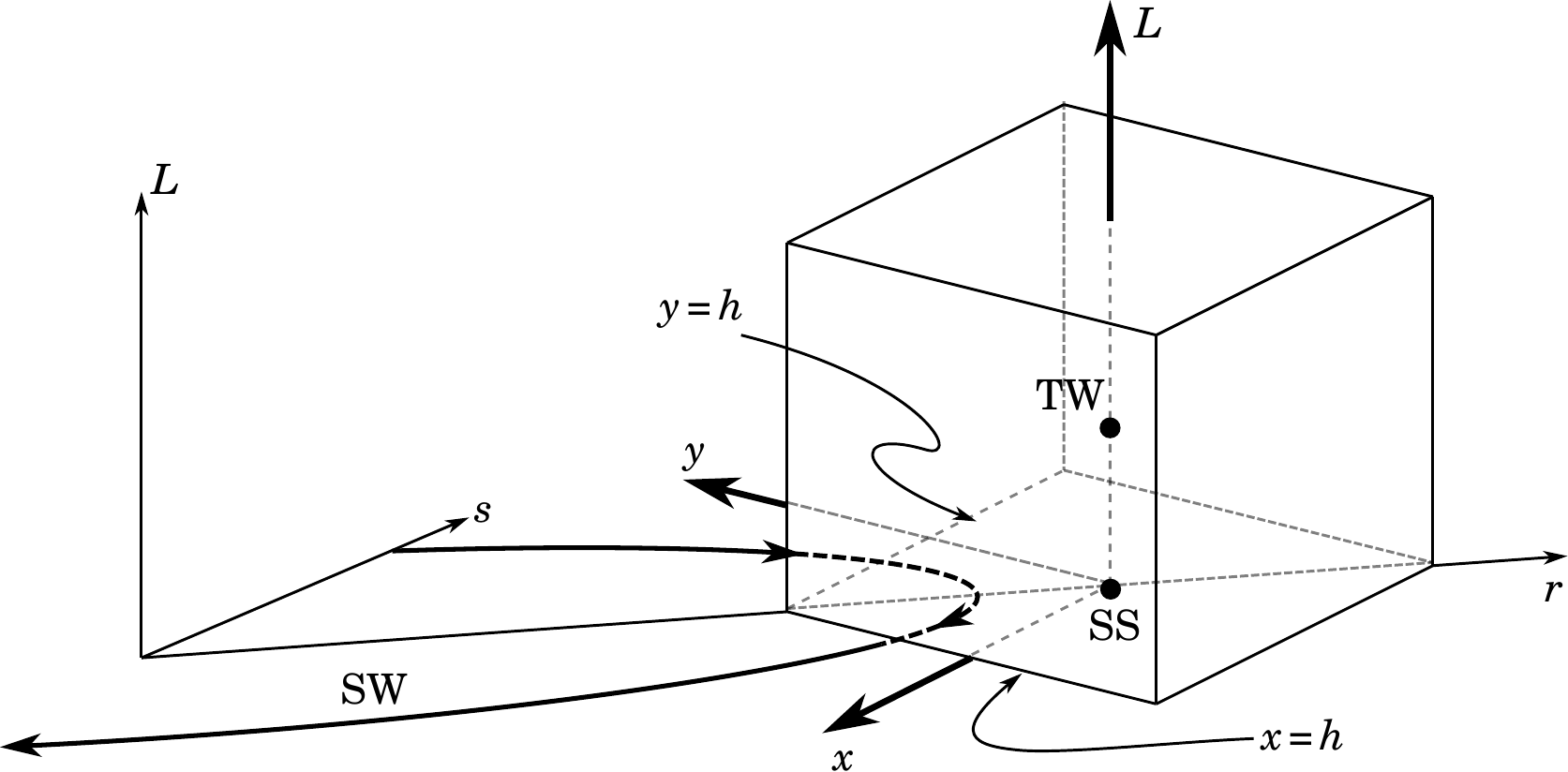,width=\hsize}}
 \end{center}
 \caption{\label{fig:poincare}Schematic diagram of the original
$(r,s,L)$ coordinate system and the new $(x,y,L)$ coordinate system
based at the $SS$ equilibrium point. The $x$ and $y$ axes point,
respectively, along the unstable and stable eigendirections
of the SS equilibrium within the $L=0$ plane. The SW periodic orbit
lies in the $L=0$ plane and 
approaches SS from the positive $y$~direction, and leaves in
the positive~$x$ direction. The illustration has $\alpha<0$ so TW exist, and 
$\kappa<0$ so SW exist. The flow defines a map from the plane $y=h$ (one of 
the back faces of the box) to itself.}
 \label{fig2Dmap}
 \end{figure}

\section{Analysis near the crossing point}
\label{secCrossing}

We derive a two-dimensional map that describes the dynamics of the third-order 
ODE~(\ref{eqnfrL}) near the 
codimen\-sion-two crossing point seen in figure~\ref{figcrossing1}(c) and in 
figure~\ref{figunfNearMminusfive}. At this point, the heteroclinic bifurcation 
where the SW periodic orbit collides with the SS equilibria coincides with 
the pitchfork bifurcation, where TW are created from~\hbox{SS}. 

For this derivation, we locate the origin of a three-dimensional coordinate 
system $(x,y,L)$ at the SS equilibrium at $(r,s,L)=(\sqrt{-\mu/A},0,0)$, 
with $x$ along the unstable manifold within the $L=0$ plane, $y$~along the 
stable manifold within the same plane, and $L$ as before (see 
figure~\ref{fig:poincare}).
We orient the coordinate system so that, close to the global bifurcation,  the
SW periodic orbit approaches SS from the positive $y$~direction, and leaves in
the positive~$x$ direction.

The derivation proceeds along standard lines, but with the complication that we
allow a pitchfork bifurcation from the SS equilibrium, cf. \cite{Hirschberg1992,Hirschberg1993,Siggers2003}.
We define a small neighborhood of the SS by $|x|\leq h$ and $|y|\leq
h$, where $h$~is small, and consider the dynamics within this neighborhood,
linearizing in the $(x,y)$ plane, but keeping nonlinear terms in the
$L$~direction:
 \begin{equation}\label{eqnlinearised}
 {\dot x} = \lambda_+ x,\qquad
 {\dot y} = \lambda_- y,\qquad
 {\dot L} = \alpha L + L^3, 
 \end{equation}
where $\lambda_+$ and $\lambda_-$ are the positive and negative eigenvalues
of~SS with eigen\-directions within the $L=0$ plane, $\alpha$ is the eigenvalue
corresponding to the~$L$ direction (so $\alpha=0$ at the pitchfork bifurcation,
when $\mu=A\nu/D$ from (\ref{eqnpfSSTW})), and we have rescaled~$L$ to set the
coefficient of the cubic term to~$+1$ (so that TW exist when $\alpha<0$). The
TW then satisfy $x=y=0$ and $L^2=-\alpha$, and the stable manifold of TW 
intersects the plane $y=h$ at $x=0$, $L^2=-\alpha$.

Trajectories enter the neighborhood with $y=h$, and initial values $x=x_0$ and 
$L=L_0$. We solve~(\ref{eqnlinearised}) to find
 \begin{equation}\label{eqnlocal}
 x(t) = x_0 e^{\lambda_+ t},\qquad
 y(t) = h e^{\lambda_- t},\qquad
 L(t) = L_0 
    \sqrt{\frac{\alpha e^{2\alpha t}}
               {\alpha + L_0^2\left(1-e^{2\alpha t}\right)}} .
 \end{equation}
Trajectories leave the neighborhood when $x=h$ at time 
$T=\frac{1}{\lambda_+}\log\left(\frac{h}{x_0}\right)$. At this time, the $y$ 
and~$L$ coordinates are:
 \begin{equation}\label{eqnlocalmap}
 y(T) = h \left(\frac{x_0}{h}\right)^{\delta_{SS}},\qquad
 L(T) = L_0 
    \sqrt{\frac{\alpha \left(x_0/h\right)^{2\delta_L}}
               {\alpha + L_0^2\left(1-\left(x_0/h\right)^{2\delta_L}\right)}},
 \end{equation}
where $\delta_{SS}=-\lambda_-/\lambda_+$ and $\delta_L=-\alpha/\lambda_+$. For
this problem, it can be shown that $\delta_{SS}$ tends to~$1$ from above in the
limit $\epsilon\rightarrow0$. 
We focus on the case where the crossing point moves 
close to the Takens--Bogdanov point as $M\rightarrow-5$ from above, and
we choose the time-scale so that $\lambda_+=1$, i.e., 
$\delta_L=-\alpha$ and suppose that $\alpha<0$ ($\delta_L>0$).

The positive branch of the unstable manifold of SS then leaves the
neighborhood at $x=h$, $y=0$ and $L=0$, and returns at $x=-\kappa$, $y=h$ and
$L=0$, where $\kappa$ is a parameter that controls how close we are to the
global bifurcation: we choose the parameter so that SW exist when $\kappa<0$,
before the global bifurcation, and that SW are destroyed in the global
bifurcation at~$\kappa=0$. The negative branch of the unstable manifold of SS
leaves the neighborhood with $x<0$ and escapes from the domain of the map.

%To see the connection between the ODE and the map parameters, in 
%figure~\ref{figunfNearMminusfive}, the line $\alpha=0$ is labeled SS-pf, and 
%the line $\kappa=0$ is labeled SW-het. The $\kappa<0$, $\alpha<0$ quadrant 
%contains the lines SW-pf, MW-pd and MW-hom.
To see the connection between the ODE and the map parameters, in 
figure~\ref{figunfNearMminusfive}, the line $\alpha=0$ is labeled SS-pf, and 
the line $\kappa=0$ is labeled SW-het. The $\kappa<0$, $\alpha<0$ quadrant 
contains the lines SW-pf, MW-pd and MW-hom. At the crossing point, where
$(\kappa,\alpha)=(0,0)$, we have $\delta_{SS}>1$. As $M\rightarrow -5$ and the
crossing point moves closer to the TB point, $\delta_{SS}\rightarrow 1^+$.

Trajectories that leave the neighborhood of SS at $(h,y(T),L(T))$, 
close to the positive branch of the unstable manifold, return to the
neighborhood at $(x',h,L')$, where, to leading order in a Taylor series
expansion, we have
 \begin{equation}\label{eqnglobalmap}
 x' = -\kappa + Ey(T) + FL^2(T),\qquad
 L' = GL(T).
 \end{equation}
The constants $E$, $F$ and $G$ are properties of the global flow, and must 
satisfy $E>0$ (since trajectories in the plane $L=0$ cannot cross) and $G>0$ 
(since trajectories cannot cross the $L=0$ subspace).
The form of this map respects the $L\rightarrow-L$ symmetry of~(\ref{eqnfrL}).

Putting this together results in a map $(x,h,L)\rightarrow(x',h,L')$, where
 \begin{equation}\label{eqnmapunscaled}
 x' = -\kappa + E x^{\delta_{SS}} + 
        \frac{F L^2 x^{2\delta_L}}
             {1+\frac{L^2}{\alpha}\left(1-x^{2\delta_L}\right)},\qquad
 L' = \frac{G L x^{\delta_L}}
            {\sqrt{1+\frac{L^2}{\alpha}\left(1-x^{2\delta_L}\right)}},
 \end{equation}
and we have dropped the $0$ subscripts and absorbed $h$ into the
definitions of the other constants and into the scaling of the $x$~coordinate. 
By further scaling~$L$ and~$\alpha$, it is possible to set $F=\pm1$. 

%However, we 
%choose not to scale $E$ to~$1$ since this scaling is valid only when
%$\delta_{SS}\neq1$~\cite{Rucklidge1993} and in our problem, $\delta_{SS}$~is
%very close to~$1$. 

Fixed points of the map~(\ref{eqnmapunscaled}) correspond to periodic orbits 
(SW and MW) in the differential equations. Within the $L=0$ subspace, we have 
SW at $(x_{SW},0)$, which satisfy:
 \begin{equation}\label{eqnSWmap}
 \kappa = - x_{SW} + E x_{SW}^{\delta_{SS}},
 \end{equation}
confirming that SW exist for $\kappa<0$ when $\delta_{SS}>1$. The stability of
SW are given by the two Floquet multipliers (eigenvalues of the Jacobian
matrix), which are $\tilde{E}=\delta_{SS} E x_{SW}^{\delta_{SS}-1}$ and
$Gx_{SW}^{\delta_L}$. Since we expect SW to be stable within the $L=0$
subspace, we deduce that $\tilde{E}$ must satisfy $0<\tilde{E}<1$, and since
this is true even in the limit $\epsilon\rightarrow0$
($\delta_{SS}\rightarrow1$), we must also have $0<E<1$. Writing $\tilde{E}$ in
this way emphasises the fact that $\tilde{E}$ is only a weak function of~$x$ when
$\delta_{SS}$ is close to~$1$, and in the limit $\epsilon\rightarrow0$
($\delta_{SS}\rightarrow1^+$),
$\tilde{E}=E$.

The second multiplier gives the location of SW-pf, where SW lose stability to
MW in a pitchfork bifurcation. Solving the equation $Gx_{SW}^{\delta_L}=1$
results in a relation between $\kappa$ and $\alpha$ for the location of SW-pf:
 \begin{equation}\label{eqnSWpfmap}
 \kappa_{\mbox{\small SW-pf}}=EG^{\delta_{SS}/\alpha} - G^{1/\alpha}.
 \end{equation}
Since SW exist for $\kappa<0$, and evidence in figure~\ref{figcrossing1}(c) and
figure~\ref{figunfNearMminusfive} suggests that the line of SW-pf connects to
the codimen\-sion-two point $(\kappa,\alpha)=(0,0)$ from the quadrant $\kappa<0$
and $\alpha<0$, we require $G>1$ for the limit of $\kappa_{\mbox{\small SW-pf}}$ to be
zero when $\alpha\rightarrow0$ from below. When $\delta_{SS}=1$, we also
require $E<1$ (as found already above).

The MW periodic orbit can be found as a fixed point of the
map~(\ref{eqnmapunscaled}) with $L\neq0$. The equation $L'=L$ implies that
$G^2x^{2\delta_L}={1+\frac{L^2}{\alpha}\left(1-x^{2\delta_L}\right)}$.
This can be solved in two equivalent ways:
 \begin{equation}\label{eqnMWexpressions}
 \begin{array}{l}
 L^2=\delta_L(1-G^2x^{2\delta_L})/(1-x^{2\delta_L}),\\
 x^{2\delta_L}=(\delta_L-L^2)/(G^2\delta_L-L^2)
 \end{array}
 \end{equation}
(recall
$\alpha=-\delta_L$). Then the equation $x'=x$ reduces to $x=-\kappa + E
x^{\delta_{SS}} + FL^2/G^2$, or $\kappa=-x+ E x^{\delta_{SS}} +
F(\delta_L/G^2)(1-G^2x^{2\delta_L})/(1-x^{2\delta_L})$. Thus, for a given value
of $x$ and the parameter $\alpha=-\delta_L$, one can calculate the
corresponding value of the parameter~$\kappa$ and of~$L^2$, for MW with this
value of~$x$. 

%The range of values of~$x$ for which MW exist is bounded by $x=G^{-1/\delta_L}$
%($L=0$), where the MW are created in SW-pf, and $x=0$ ($L^2=\delta_L$), where
%they are destroyed in a homoclinic bifurcation (MW-hom) when they collide with
%the stable manifold of~\hbox{TW}. 
%The upper bound of this range goes to zero as $\delta_L\rightarrow0^+$.
%At the MW-hom bifurcation, we have 
The range of values of~$x$ for which MW exist is bounded by $x=G^{-1/\delta_L}$
($L=0$), where the MW are created in SW-pf, and $x=0$ ($L^2=\delta_L$), where
they are destroyed in a homoclinic bifurcation (MW-hom) when they collide with
the stable manifold of~\hbox{TW}. Since $G>1$ the upper bound of this range
decreases to zero as $\delta_L\rightarrow0^+$. At the MW-hom bifurcation, we have 
 \begin{equation}\label{eqnMWhom}
 \kappa_{\mbox{\small MW-hom}}=-\alpha F/G^2. 
 \end{equation}
Numerical evidence in figure~\ref{figcrossing1}(c) and
figure~\ref{figunfNearMminusfive} suggests that this line is also in the
$\kappa<0$ and $\alpha<0$ quadrant, and we conclude that $F=-1$.

With this value of $F$ the Jacobian matrix evaluated on the MW branch is:
 \begin{equation}\label{eqnJacobianMW}
 J=\left[
   \begin{array}{cc}
   \tilde{E}
      - \frac{2L^2\delta_L}{G^2x}
      + \frac{2L^4}{G^4 x} &
   \frac{-2L}{G^4 x^{2\delta_L}} \\
   \frac{L\delta_L}{x} - \frac{L^3}{G^2x} &
   \frac{1}{G^2 x^{2\delta_L}}
   \end{array}
   \right],
\quad\mbox{with}\quad 
 \mbox{det}(J)=\frac{\tilde{E}}{G^2x^{2\delta_L}},
 \end{equation}
where $x$ and $L$ are the values for the MW, and we have made use
of~(\ref{eqnMWexpressions}) to make these expressions as simple as
possible. In these expressions, $\tilde{E}=\delta_{SS} E x^{\delta_{SS}-1}$.
Since $\delta_{SS}>1$ and $\delta_L=0$ at the crossing point, it follows that
${\rm det}(J)\rightarrow 0$ at this point.
%where $x$ and $L$ are the values for the MW, and 
%we have made use of~(\ref{eqnMWexpressions}) to make these expressions as 
%simple as possible. In these expressions, 
%$\tilde{E}=\delta_{SS} E x^{\delta_{SS}-1}$.

We consider first the possibility of a torus bifurcation from MW; the relevant 
conditions are $\mbox{det}(J)=1$ and $-2<\mbox{Trace}(J)<2$. The first equation
implies $x^{\delta_L}=\sqrt{\tilde E}/G$. With this value of $x$, we have
$L^2=\delta_LG^2(1-\tilde{E})/(G^2-\tilde{E})$, so
%We consider first the possibility of a torus bifurcation from MW; the relevant 
%conditions are $\mbox{det}(J)=1$ and $-2<\mbox{Trace}(J)<2$. The equation
%implies $x^{\delta_L}=\sqrt{\tilde E}/G$. With this value of $x$, we have
%$L^2=\delta_LG^2(1-\tilde{E})/(G^2-\tilde{E})$, so
 \begin{equation}
 \mbox{Trace}(J) = \tilde{E} + \frac{1}{\tilde{E}} 
                   - \frac{2(G^2-1)(1-\tilde{E})\delta_L^2}
                          {(G^2-\tilde{E})^2
                       \left(\frac{\sqrt{\tilde{E}}}{G}\right)^{1/\delta_L}}.
 \end{equation}
For small positive~$\delta_L$ (close to the codimension-two crossing point),
we have $\delta_L^2/(\sqrt{\tilde{E}}/G)^{1/\delta_L}\rightarrow\infty$, so 
the condition $-2<\mbox{Trace}(J)<2$ is not satisfied, and there is no torus 
bifurcation from MW connecting to the codimension-two point.

We turn now to the period-doubling case. A numerical exploration of the
relevant condition, $\mbox{det}(J) + \mbox{Trace}(J) + 1=0$, suggests that
with small~$\delta_L$, there are solutions close to the SW-pf bifurcation,
with $Gx^{\delta_L}$ close to~$1$ and $L$ close to~$0$ (in particular, $L^2\ll\delta_L$).
With this in mind, we write the condition $\mbox{det}(J) + \mbox{Trace}(J) + 1=0$ as
%We turn now to the period-doubling case. A numerical exploration of the
%relevant condition, $\mbox{det}(J) + \mbox{Trace}(J) + 1=0$, suggests that with
%small~$\delta_L$, there are solutions close to the SW-pf bifurcation, with
%$Gx^{\delta_L}$ close to~$1$ and $L$ close to~$0$ (in particular,
%$L^2\ll\delta_L$). With this in mind, we write the condition[6~[6~ as
 \begin{equation}\label{eqnPDMW}
 \frac{(L^2+G^2L^2 - 2\delta_LG^2)(1+\tilde{E})}
      {L^2-\delta_L}
 =
 \frac{2(\delta_L G^2-L^2)L^2}
      {G^2x}.
 \end{equation}
With $x\approx(1/G)^{1/\delta_L}$ and $L^2\ll\delta_L$, this is solved by
 \begin{equation}
 L^2 \approx \frac{1}{\delta_L}\left(\frac{1}{G}\right)^{\frac{1}{\delta_L}} 
             G^2(1+\tilde{E}),
 \end{equation} 
from which $\kappa$ can be deduced:
 \begin{equation}\label{eqnPDMWkappa}
 \kappa_{\mbox{\small MW-pd}} =
 -\left(1+\frac{1}{\delta_L}\right)
         G^{\frac{1}{\alpha}}
 + E G^{\frac{\delta_{SS}}{\alpha}}
     \left(1-\frac{\delta_{SS}}{\delta_L}\right)
 \end{equation}
To obtain this expression we have replaced $\delta_L$ by $-\alpha$.
This approximate solution satisfies the scaling 
assumptions that were made in its derivation.
There is therefore a line of MW-pd bifurcations connecting to the codimension-two 
crossing point. Comparing with~(\ref{eqnSWpfmap}), we see that the bifurcation 
lines SW-het (creating SW at $\kappa=0$), SW-pf (creating the MW) and MW-pd
(where the MW lose stability) are all tangent as they approach the
codimension-two crossing point. These results have been used to inform the sketch
in figure~\ref{figunfNearMminusfive}.

As $M\rightarrow-5$ (and $\delta_{SS}\rightarrow 1^+$), the crossing point
approaches the TB point and brings with it the MW-pd bifurcation found above.
Consequently, when $M=-5$, we expect to see
the MW-pd bifurcation line emerging from the TB point; this is supported
by the numerical evidence in figure~\ref{figcrossing2}(a). The map does not 
apply for $M<-5$ (since there is no crossing point), but numerical 
evidence (figure~\ref{figcrossing2}b,c) indicates that the MW-pd line remains 
connected to the TB point.

%again using $\delta_L=-\alpha$. This approximate solution satisfies the scaling 
%assumptions that were made in its derivation.
%There is therefore a line of MW-pd bifurcations connecting to the codimension-two 
%crossing point. Comparing with~(\ref{eqnSWpfmap}), we see that the bifurcation 
%lines SW-het (creating SW at $\kappa=0$), SW-pf (creating the MW) and MW-pd
%(where the MW lose stability) are all tangent as they approach the
%codimension-two crossing point.
%These results have been used to inform the sketch in 
%figure~\ref{figunfNearMminusfive}.

%As $M\rightarrow-5$, the crossing point approaches the TB point and brings 
%with it the MW-pd bifurcation found above, so when $M=-5$, we expect to see
%the MW-pd bifurcation line emerging from the TB point; this is supported
%by the numerical evidence in figure~\ref{figcrossing2}(a). The map does not 
%apply for $M<-5$ (since there is no crossing point), but numerical 
%evidence (figure~\ref{figcrossing2}b,c) indicates that the MW-pd line remains 
%connected to the TB point.

\section{Discussion}
\label{secDiscussion}

The presence of chaos near the Takens--Bogdanov point has been of interest ever
since the discovery of chaotic dynamics of SW solutions in doubly diffusive
convection~\cite{Moore1983,Knobloch1986a} and its study established its
relation to the Shil'nikov scenario, i.e., to the presence of a Shil'nikov
orbit between a pair of symmetry-related saddle-points corresponding to
unstable steady convection, all within the $L=0$ invariant subspace of the
governing PDEs
\cite{Knobloch1986a,Proctor1990,Rucklidge1992,Rucklidge1993,Rucklidge1994}.
However, it was discovered in \cite{Knobloch1986f} that the SW states are in
fact unstable to TW perturbations whenever such waves are permitted by the
imposed boundary conditions, a discovery that led to intensive study of the
competition between SW and TW in the nonlinear regime in systems with periodic
boundary conditions \cite{Knobloch1986c,CLUNE1992,Matthews1993}. The study of the Takens--Bogdanov
bifurcation with $O(2)$ symmetry was the natural next step, undertaken in DK,
since this bifurcation brings steady states into the picture as well. The
results of the DK analysis were subsequently applied to a number of systems
including convection in a magnetic field \cite{Dangelmayr1986a,Knobloch1986e}
and rotating convection \cite{Knobloch1990}.

However, as shown here, the analysis of DK is incomplete in one important
sense, associated with the failure of averaging near global 
bifurcations.\footnote{DK also omit one of the bifurcation diagrams that arises
in region IX$^-$ of their classification \cite{Peplowski1988}.} This difficulty
has in fact been known for a long time \cite{Guckenheimer1984}. In particular, 
it was known that the integrability of many normal forms gives incomplete 
results in regions near global bifurcations since either higher order terms or 
those that break normal form symmetry can lead to transverse intersection of 
stable and unstable manifolds and these in turn generate thin regions of chaos 
near these manifolds. This is the case in the Takens--Bogdanov bifurcation with
$O(2)$ symmetry as well, and in particular in the vicinity of the global
bifurcation at which the MW branch terminates. The reason is simple: averaging
requires that the oscillation period (here the modulation period) remains
finite for small $\epsilon$, a condition that breaks down near global
bifurcations. In these regions a higher-dimensional analysis is required, and
as shown here, period-doubling and chaos may be uncovered. When the homoclinic
or heteroclinic tangles are generated by terms breaking normal form symmetry
these parameter regions are exponentially thin; this is expected to be the case 
in the present study as well, and accounts for the difficulty in carrying out 
numerical work for small~$\epsilon$.

Of particular interest in the present case is the fact that some of the
resulting complex behavior is present arbitrarily close to the codimension-two
Takens--Bogdanov point, i.e., arbitrarily close to the bifurcation. We consider
this bifurcation, therefore, to be another example of instant chaos, much as
the well-known cases studied by Arneodo \etal \cite{Arneodo1985} and Guckenheimer 
and Worfolk \cite{Guckenheimer1992a}. In contrast the SW chaos referred to above 
cannot take place arbitrarily close to the TB point. In the
2:1 spatial resonance with $O(2)$ symmetry the heteroclinic cycles present near
the codimension-two point are structurally stable \cite{Armbruster1988} and any
chaos in this regime is the result of the presence of noise. However, further
out from the codimension-two point one finds the same transition scenario as
found here, viz. SS$\to$TW$\to$MW followed by a global bifurcation involving SW
\cite{Porter2001}.

The fact that the full, unaveraged yet truncated normal form does in fact
capture some aspects of the chaotic dynamics associated with the termination of
MW implies that the normal form contains information beyond that in the
averaged equations studied by DK. Evidently it is not necessary to go to higher
order in the normal form in order to capture the behavior observed in
\cite{Knobloch1990b,Knobloch1991,Cox1992}, only to extend the normal form
analysis beyond averaging. In contrast, in other normal forms exhibiting global
bifurcations \cite{Hirschberg1993,Siggers2003,Kirk2004,Simpson2005} the normal
forms are intrinsically two-dimensional, at least after symmetry reduction. In
these cases complex dynamics appear only when the normal form symmetry is
explicitly broken.  However, as cautioned by Wittenberg and Holmes \cite{Wittenberg1997},
the resulting interval of chaos may be much thinner than in the original system
from which the normal form was derived.

The fact that averaging fails near global bifurcations such as those discussed here
involving the termination of the MW branch is of course well known. This failure originates
in the divergence of the oscillation period at the global bifurcation or, as in the present
case, the divergence of the modulation period. A common procedure for overcoming this problem 
is to construct an appropriate return map valid in the neighborhood of this bifurcation and
matching the results to those obtained from averaging as one moves away from the global
bifurcation. This is the procedure adopted here. In \cite{Hirschberg1993} the authors
consider a fixed point undergoing a Hopf and a pitchfork bifurcation in close succession
in a regime where the mixed mode undergoes a secondary Hopf bifurcation, thereby creating
an invariant 2-torus. The authors show that terms that break normal form symmetry generate 
a sequence of resonances between the primary and secondary Hopf frequencies and mergers of
the resonance tongues \cite{Kirk1993} may result in complex dynamics of Shil'nikov type associated 
with homoclinic bifurcations that are intertwined with heteroclinic bifurcation. This complex
structure results from the break up of the heteroclinic orbit responsible for the destruction
of the 2-torus in the normal form. In the problem studied here we find related behavior,
including Shil'nikov dynamics. The reason is fundamentally the same: the original system (\ref{eqnfrL})
is three-dimensional, but the process of averaging leads to a second order system and hence
captures only two of the three eigenvalues required to specify the stability properties
of the MW 2-torus completely. These two eigenvalues describe these properties correctly
when the averaging procedure is valid because the third eigenvalue vanishes. But in the
vicinity of the global bifurcation the third eigenvalue no longer decouples and its sign
determines the stability of the MW near its termination. Indeed, we can view the results 
of this paper as showing that this eigenvalue is in fact negative (so that the MW branch
is unstable near its end), despite the fact that the averaged system studied in \cite{Dangelmayr1987b}
suggests that it should be stable, based on the two eigenvalues that can be computed
using this approach. It is this realization that reconciles the results of the present
work and that of DK. In other systems this increase in the dimension of the system near
global bifurcations leads to thin regions of chaos, as described, for example, by 
Guckenheimer \cite{Guckenheimer1981,Guckenheimer1984} and Langford \cite{Langford1983}.

%Discuss problems of averaging, and what is does to the FMs. Refs (see above, 
%and we scribbled down a few notes: Guckenheimer 1980/81; Kirk: cusp-shaped 
%region of heteroclinic tangles; Iooss). What do DK say about the stability of 
%the MW at the homoclinicity~\cite[p.~263]{Dangelmayr1987b}?

\section*{Acknowledgments} 
 % \begin{ack}
 
This work was begun in 1993. We thank Marty Golubitsky for encouraging us to 
publish the results in a timely fashion.

 % \end{ack}

%More references to include? \cite{Knobloch1991}

% ------------------------------------------                                    

\begingroup

\def\url#1{}
\def\issn#1{}
\def\issnp#1{}

% this is for bibtex

% \bibliographystyle{plain} 
% \bibliographystyle{cDSS} 
\bibliographystyle{elsart-num}

\bibliography{rk}

\endgroup

% ------------------------------------------                                    

% \begin{thebibliography}{99}
% 
% \bibitem{Chandrasekhar1961} 
% S. Chandrasekhar
% \newblock Hydrodynamic and Hydromagnetic Stability
% \newblock Clarendon Press: Oxford 1961.
% 
% \bibitem{Dangelmayr1987b} 
% G.~Dangelmayr and E.~Knobloch.
% \newblock The Takens--Bogdanov bifurcation with $O(2)$-symmetry
% \newblock {\em Phil.\ Trans.\ R.~Soc.\ Lond.~A}
%           {\bf 117} (1987) 243--279.
% 
% \end{thebibliography}
   
\end{document}